\documentclass[10pt,english,sigconf,authoryear,nonacm]{article}
\usepackage[T1]{fontenc}
\usepackage[latin9]{inputenc}
\usepackage{geometry}
\usepackage{babel}
\usepackage{amsmath}
\usepackage[unicode=true]
 {hyperref}

\makeatletter
\newcommand{\lyxaddress}[1]{
	\par {\raggedright #1
	\vspace{1.4em}
	\noindent\par}
}

\AtBeginDocument{
  
}

\makeatother

\begin{document}
\title{Danish National Election 2022 Twitter Data on Likes, Retweets, and
Botscores for the purpose of exploring Coordinated Inauthenthic Behavior}
\author{Laura Jahn$^{1}$ and Rasmus K. Rendsvig$^{2}$}
\maketitle

\lyxaddress{1,2: Center for Information and Bubble Studies, University of Copenhagen\\
1: \href{mailto:laurajahn@outlook.de}{laurajahn@outlook.de}, ORCID:
\href{https://orcid.org/0000-0002-5475-8496}{0000-0002-5475-8496}\\
2: \href{mailto:rendsvig@gmail.com}{rendsvig@gmail.com}, ORCID: \href{https://orcid.org/0000-0002-5475-8496}{0000-0002-5475-8496}}

\noindent This note primarily contains the README.md from the \href{https://github.com/LJ-9/Danish-Election-2022-Twitter-Likes-Retweets-Botscores-Inauthentic-Coordinated-Behavior}{GitHub repository}
of the same name, with a few additional comments and references. We
upload the note for visibility, hoping that other researchers may
find the data of use.

The repository contains code related to the dataset on the Danish
National Election 2022, available at \href{https://doi.org/10.7910/DVN/RWPZUN}{Harvard Dataverse}.
See the directory \emph{Data from Danish Twitter on National Election
2022} in the Harvard repository.

We cluster Twitter users into bins of users that showed exactly the
same liking/retweeting behavior over the period. To investigate whether
any of these bins exhibited \emph{coordinated inauthentic behavior},
we were interested in whether bin size correlated with
\begin{itemize}
\item user account deletion/suspension (we bought some likes at some point,
and saw that the user accounts disappeared rather quickly)
\item high bot scores from Botometer / Botometer Lite.
\end{itemize}
We didn't find significant correlations. Neither, and somewhat surprisingly,
between Botometer and Botometer Lite scores.

\subsection*{Dates for Data Collection}
\begin{itemize}
\item November 1, 2022: Election date.
\item October 7, 2022 and 30 days forward: we scraped with the query \emph{\#dkpol
-is:retweet OR \#fv22 -is:retweet OR \#fv2022 -is:retweet}. See \texttt{Pull-All\_fv22/parameters.py}.
\item Late January to Mid February 2023: we looked up user account information
and botscores.
\end{itemize}
Botometer guidelines suggest that we should have looked up botscores
runningly as they are somewhat time sensitive. Checking for correlations
with botscores, however, came as an afterthought to the general data
collection.\medskip{}

\subsection*{Code on Dataset}

To use the code in the \href{https://github.com/LJ-9/Danish-Election-2022-Twitter-Likes-Retweets-Botscores-Inauthentic-Coordinated-Behavior}{GitHub repository}
on the dataset, clone the repo and download the dataset, extract \texttt{Data
from the directory Data from Danish Twitter on National Election 2022/Raw
Data/Pull-All\_fv22.zip} and place the \texttt{Pull-All\_fv22} directory
in the repo root.\medskip{}

\subsection*{Much more Information}

For much more information on the research motivation for creating
this dataset and undertaking the analysis, please see the introduction
to the \href{https://github.com/LJ-9/Danish-Election-2022-Twitter-Likes-Retweets-Botscores-Inauthentic-Coordinated-Behavior/blob/main/documents/Jahn_Laura_PhD_Thesis_2023_online.pdf}{PhD thesis} \emph{Curbing Amplification Online: Towards Improving the Quality of Information Spread on Social Media Using Agent-Based Models and Twitter Data}
by Laura Jahn, University of Copenhagen, 2023 \cite{LauraThesis}.
A related paper extensively discussing data collection and preliminary
analysis on the same type of data (likes) can be found on \href{https://arxiv.org/abs/2305.07384}{arXiv}
\cite{JahnRendsvigTwitterLikesCIB} and the corresponding \href{https://github.com/humanplayer2/get-twitter-likers-data}{GitHub repository}
\cite{JahnRendsvig22GetLikers}.

\section*{Dataset components}

The dataset contains the following:

\subsection*{1. /Raw Data/Pull-All\_fv22.zip}

\subsubsection*{Tweet IDs, Liking Users and Retweeting Users live scraped from Oct.
7th and 30 days forward}

Runningly, we collected the user identities of liking and retweeting
users, with the algorithm described in \cite{JahnRendsvigTwitterLikesCIB}
using the code from \href{https://github.com/humanplayer2/get-twitter-likers-data}{this repo}.

The dataset contains the collected data, time stamped. I.e., for every
\textasciitilde 5 minutes in the scrape period, there is a file of
the last 48 hours' tweet IDs, and lists of liking and retweeting users.
See the linked-to repo for information on the directory and file structure.

This repo contains code to process these rather raw files. In addition
to this readme, you can also go through the file \texttt{analysis/Pull-fv22-treatment.py}
for a bit of practical code.\medskip{}

\emph{Note:} The dataset does not contain e.g. tweet text, as this
cannot to shared in batch without violating Twitter's terms and conditions.

\section*{2. /Preprocessed Data}

\emph{Note:} We have uploaded the processed data as it can take quite
some RAM to conclude the processing. We ran out of memory on 128 GB
RAM Linux machine, so we supplemented it with an additional 2 TB NVMe
disk, allocated to swap space. One treatment ended up using \textasciitilde 590
GB of that.

\subsection*{2.0 /Preprocessed Data}
Preprocessed using code from this repo, see below.

The \texttt{binarymatrices.zip} file contains two Tweet IDs $\times$
User IDs matrices (one for likes, one for retweets), with a $1$ in
$(i,j)$ if user $j$ liked/retweeted tweet $i$, else $0$.
\begin{itemize}
\item Processed using \texttt{binarymatrices.sh}
\end{itemize}

\subsection*{2.1 /Preprocessed Data/Botscores}

Contains botscores for every liking user / retweeting user observed
during the live scrape, according to \href{https://cnets.indiana.edu/blog/2020/09/01/botometer-v4/}{Botometer v4}
and \href{https://botometer.osome.iu.edu/botometerlite}{Botometer Lite}.

Collected using \texttt{botscores\_v4.sh} and \texttt{botscores\_lite.sh}

\subsection*{2.2 /Preprocessed Data/Clusters}

Contains analyses of the binary matrices, where users have been grouped
if they share the exact same liking / retweeting behavior. As described
in \cite{JahnRendsvigTwitterLikesCIB}.

The clustering/binning was done using \texttt{/analysis/binning.sh}.

\subsection*{2.3 /Preprocessed Data/Later Users and Tweets Lookups}
\begin{itemize}
\item \texttt{latercounts.pkl}: look up of all tweets to get their like
and retweet counts,
\item \texttt{laterusers.pkl}: look up of all user profiles e.g. whether
it still exists, has been suspended, error codes, etc.
\end{itemize}

\subsection*{2.4 /Preprocessed Data/Likers Retweeters Pagination}

To improve the live scraped data, we have since used Twitter's updated
API allowing for pagination to re-collect the liking and retweeting
users.

This data should, if no users were deleted or unliked, contain the
same information as the live scraped data.

See the section \emph{Disclaimer for Article II} in the \href{https://github.com/LJ-9/Danish-Election-2022-Twitter-Likes-Retweets-Botscores-Inauthentic-Coordinated-Behavior/blob/main/documents/Jahn_Laura_PhD_Thesis_2023_online.pdf}{PhD thesis}
\cite{LauraThesis} for more information.

This paginated data was collected using the \href{https://twarc-project.readthedocs.io/en/latest/twarc2_en_us/}{twarc2 package},
using the scripts \texttt{analysis/twarc-lookup-liking-users.sh} and
\texttt{analysis/twarc-lookup-retweeting-users.sh}. As we had multiple
bearer tokens available, we split the user lists using \texttt{twarc-split-user-list.py}
into sublists so we could collect data in parallel.

\section*{Testing for Correlations: Conclusions}

Again, we were mostly interested in whether the correlation in liking/retweeting
behavior among users correlated with
\begin{itemize}
\item user account deletion/suspension, creation date, and/or
\item high bot scores from Botometer / Botometer lite.
\end{itemize}
We did not find any significant correlations: The size of a user's
cluster of identically liking/retweeting users did not correlate with
the user's Botometer scores or Botometer Lite scores, and the size
of a cluster did not correlate with how many of its users had disappeared.
Also, Botometer and Botometer Lite scores did not correlate with one
another.

\subsection*{To replicate:}

1. Run \texttt{analysis/features-frame.py}: Collects tweet and user
information, including botscores, user cluster sizes, deletions, etc.

In a bit more detail, the feature frames we created to check for correlated
features included the following:
\begin{itemize}
\item binsize: size of bin/cluster the user is grouped into
\item user-ID
\item error\_title: result from user look-ups
\item error\_detail
\item name
\item created\_at: creation date of user account
\item protected: whether user account is private
\item verified: whether user account is verified
\item followers\_count: number of followers of user
\item following\_count: number of accounts the user follows
\item tweet\_count: number of posted tweets by user
\item listed\_count: number of public lists that user is a member of
\item screenname - Botscores
\item All 8 raw, universal Botometer v4 botscores, such as {*}overall{*},
{*}fake follower{*}, {*}astroturf{*}
\item Botometer Lite score
\end{itemize}
2. Run \texttt{analysis/correlation-coeff.py}: Loads the feature frame
and inspects data, e.g.:
\begin{itemize}
\item slices dataframe to filter for users in large bins,
\item computes the Pearson correlation coefficient across all columns,
\item plots histograms comparing users' different features, e.g. bin size
and bot scores.
\end{itemize}

\end{document}